\documentclass{pasj00}
\draft

\begin{document}
\SetRunningHead{Horiuchi et al.}{Milli-parsec--scale structure of Cen A}
\Received{2005/07/11}%{yyyy/mm/dd}
\Accepted{2005/08/16}%{yyyy/mm/dd}

\title{Ten milliparsec-scale structure of the nucleus region in Centaurus A}
%%%% begin:list of authors
 \author{%
 Shinji \textsc{Horiuchi},\altaffilmark{1}
%\author{
 David L. \textsc{Meier},\altaffilmark{2} \\
%\author{
 Robert A.\textsc{Preston},\altaffilmark{2}
%\author{
 Steven J.\textsc{Tingay}\altaffilmark{1}  }

%  \thanks{Example: Present Address is xxxxxxxxxx}}
 \altaffiltext{1}{Centre for Astrophysics and Supercomputing,
  Swinburne University of Technology, \\
  Mail No H39
  P.O. Box 218, Hawthorn, Vic. 3122, Australia  }
\email{E-mail(SH): shoriuchi@swin.edu.au}

 \altaffiltext{2}{Jet Propulsion Laboratory, Mail Code 238-322,
  California Institute of Technology, \\
  4800 Oak Grove Drive,  Pasadena, CA 91109  }

%\author{B-Firstname \textsc{B-Familyname}}
%\affil{B-Address of Institute}\email{bbbbb@xxx.xxx.xx.xx}
%\and
%\author{C-Firstname {\sc C-Familyname}}
%\affil{C-Address of Institute}\email{ccccc@xxx.xxx.xx.xx}
%%%%end:list of authors

%%% Please use the following style in case that sorting by 
%%% affilation is impossible. 
%
% \author{%
%   D-Firstname \textsc{D-Familyname}\altaffilmark{1}
%   E-Firstname \textsc{E-Familyname}\altaffilmark{1,2}
%   and
%   F-Firstname \textsc{F-Familyname}\altaffilmark{2}}
% \altaffiltext{1}{Address of Institute}
% \email{ddddd@xxx.xxx.xx.xx}
% \email{eeeee@xxx.xxx.xx.xx}
% \altaffiltext{2}{Address of Institute}

%% `\KeyWords{}' always has to be placed before `\maketitle'.
 %Do NOT move this preamble from here!

\maketitle

\begin{abstract}
  
  We present the results of a VLBI Space Observatory Programme (VSOP)
  observation of the subparsec structure in Centaurus A at 4.9 GHz.
  The observation produced an image of the sub-parsec jet components
  with a resolution more than 3 times better than images from previous
  VLBI monitoring campaigns at 8.4 GHz and more than 2 times better
  than the previous 22 GHz studies.  Owing to its proximity, our
  Centaurus A space-VLBI image is one of the highest spatial
  resolution images of an AGN ever made --- 0.01 pc per beam ---
  comparable only to recent 43 GHz VLBI images of M87. The elongated
  core region is resolved into several components over 10
  milli-arcseconds long (0.2 pc) including a compact component of
  brightness temperature 2.2 $\times 10^{10}$K.  The counter jet was
  detected: if we assume jet--counterjet symmetry, a relatively slow
  jet speed, and a large viewing angle, as derived from previous
  observations, the image allows us to investigate the distribution of
  ionized gas around the core, which is opaque at this frequency due
  to free-free absorption.  We also analyze the jet geometry in terms
  of collimation.  Assuming the strongest component to be the core,
  the jet opening angle at $\sim 5000r_s$ from the core is estimated
  to be $\sim 12^\circ$, 
  %with collimation of the jet occurring and
  with collimation of the jet to $\sim 3^\circ$ continuing out to 
  $\sim 20,000r_s$.
  %continuing out to the lower end of the jet.  
  This result is consistent with previous studies of the jet in M87, 
  which favor MHD disk outflow models. Future space VLBI observations 
  at higher frequencies will probably be able to image the collimation 
  region, within 1000$r_s$ of the center of Centaurus~A, together 
  with the accretion disk itself.

\end{abstract}

\section{Introduction}

Centaurus A (NGC 5128) is the closest radio galaxy, at a distant 
of only 3.4 Mpc (Israel 1998). 
This proximity allows
higher resolution imaging than can be obtained for more distant galaxies,
resulting in
Centaurus A being one of the most important targets for the study of jet 
formation and collimation mechanisms in active galactic nuclei (AGNs). 
The estimated mass of the central black hole, based on infrared observation
of a 20pc central disk, is 2.0$\pm_{1.4}^{3.0} \times 10^8M_\odot$
(Marconi et al. 2001).
Observations with VLBI arrays can obtain images of the nuclear radio
source in this galaxy with a linear resolution of 0.02 pc ($\sim 1000r_s$)
for a 1 mas angular resolution.
 
Previous VLBI monitoring of Centaurus A shows that the subparsec-scale
jet has components with a slow proper motion away from the nucleus of
approximately 0.1c, while irregular episodes of rapid evolution within
jet component suggests that the underlying flow of the jet is much
faster, and probably greater than 0.45c (\cite{key-3}, \cite{key-4}).
Considering the likely jet speed and the jet-to-counterjet surface
brightness ratio (\cite{key-2}), Tingay et al.\ (1998) concluded that
the sub-parsec scale jet is inclined to our line of sight by between
50$^\circ$ and 80$^\circ$.  These results are in contrast with a
recent analysis based on the VLA observations for 100pc scale jet,
which suggest apparent subluminal motions of approximately 0.5$c$ and
a significantly smaller angle to the line of sight of $\sim$
15$^\circ$ (Hardcastle et al. 2003).
  Using the VLBA at 43 GHz, Kellermann et al. (1997) measured the size 
  of the compact radio nucleus of Centaurus A to be about 0.5 mas (0.01 
  pc) by fitting the visibilities with a circular Gaussian component. 
  However the coverage of (u,v) plane with the VLBA alone is severely 
  limited because the source is so far south, and they were not able to 
  construct an image of the nucleus.

To study the nature of nuclear region, we conducted a space VLBI (VSOP) 
monitoring program of Centaurus A, observing with the HALCA space telescope
(Hirabayashi et al. 1998) at 4.9 GHz and 1.6 GHz, from mid 1999 
to mid 2000 when the ($u,v$)-coverage was optimal
(Horiuchi, Meier and Preston 2002). 
Space-ground baseline fringes of Centaurus A were detected for the first 
time, although the signal to noise ratio became rapidly smaller as 
the space baseline length became larger than one Earth diameter. 
These observations produced images of the sub-parsec jet components 
with a resolution several times better than images in an ongoing 8.4 GHz 
monitoring campaign and a few times better than 22 GHz studies. 

%An analysis based on the multi-epoch data is presented
%elsewhere (Horiuchi, Meier and Preston 2002, Horiuchi et al. in preparation). 
In this paper we highlight the highest quality image obtained, 
%which is from the last experiment in the series at the epoch of 
from the observation on 
2000 August 6, due to sensitive telescopes, especially 
the phased VLA and ATCA, which joined the ground telescope array.
Owing to its proximity, our Centaurus A space-VLBI images are 
some of the highest spatial resolution images ever made of an AGN
---  0.01 pc per beam  --- comparable only 
to the 43 GHz VLBI images of M87 (Junor et al. 1999). 
We discuss the structure of the core region and the implications of 
the remarkably collinear jet extending from the core region to the
1 pc scale. 
%%the complex behavior of the sub-parsec jet flow.

%\newpage

\section{Observations and Results}

%In figure \ref{fig:sample}, ...

\begin{figure}
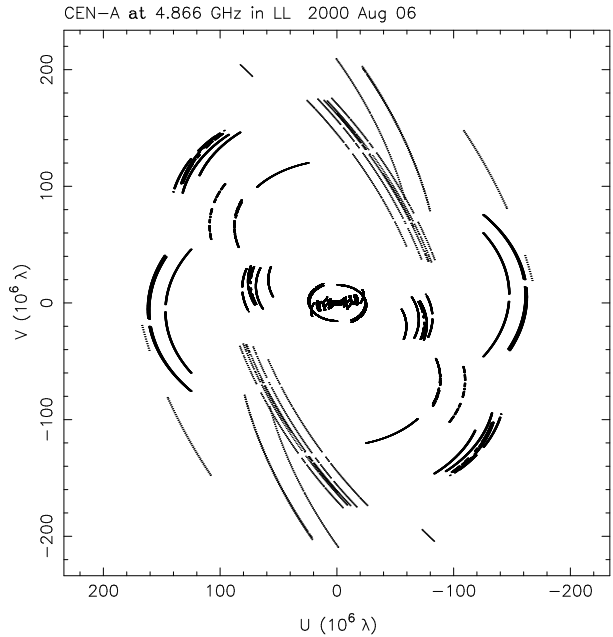

  \begin{center}
%    \FigureFile(80mm,80mm){CENA90g_gs_uvpl_v.ps}
    \FigureFile(80mm,80mm){figure1.ps}
    %%% \FigureFile(width,height){filename}
  \end{center}
  \caption{The ($u,v$)-coverage of the VSOP observation. 
Solid lines are for ground 
baselines and dotted lines 
(the long, straight diagonal tracks)
are space baselines where fringe-fitting 
was successful.}\label{fig:sample}
\end{figure}

\begin{figure}
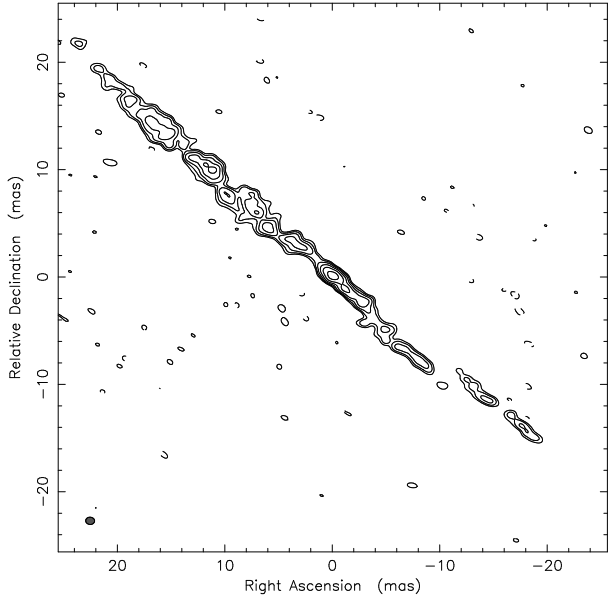

  \begin{center}
%    \FigureFile(80mm,80mm){CENA90g_gs_zoomzm_test.ps}
    \FigureFile(80mm,80mm){figure2.ps}
    %%% \FigureFile(width,height){filename}
  \end{center}
  \caption{The full resolution VSOP image of the inner 40 mas around the peak. 
The FWHM of the restored beam is 0.827 $\times$ 0.68 mas at a P.A. 
of 86.2$^\circ$ as indicated in the small filled circle in the bottom left hand corner.
Contours are drawn at -2, 2, 4, 8, 16, 32 and 64\% of 0.182 Jy/beam, 
the peak flux density in the map.}\label{fig:sample}
\end{figure}

\begin{figure}
  \begin{center}
%    \FigureFile(200mm,200mm){CENA90g_label.eps}
%    \FigureFile(200mm,200mm){CENA90g_pa_revimg.eps}
%    \FigureFile(200mm,200mm){figure3.eps}
    \FigureFile(200mm,200mm){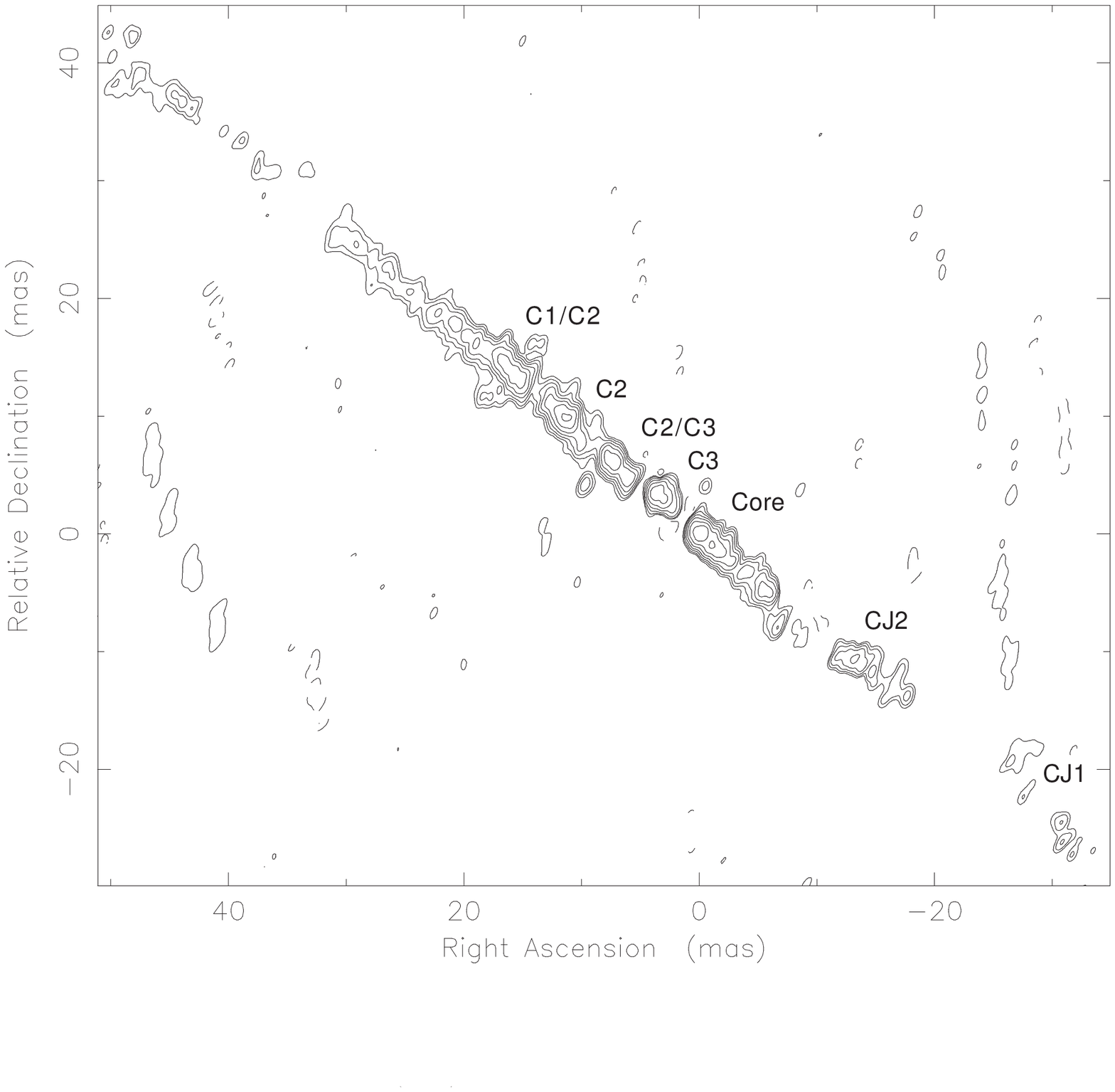}
    %%% \FigureFile(width,height){filename}
  \end{center}
  \caption{The image obtained from the ground telescopes only.
The FWHM of the restored beam is 1.19 $\times$ 0.61 mas at P.A. -27.5$^\circ$.
Contours are drawn at 
%-1, 1, 2, 4, 8, 16, 32 and 64\% 
-0.8, 0.8, 1.6, 3.2, 6.4, 12.8, 25.6 and 51.2\% 
of 0.167 Jy/beam, 
the peak flux density in the map.}\label{fig:sample}
\end{figure}

The observations presented here were conducted as a part of VSOP observation 
series to monitor the evolution of Centaurus A over 18 months starting
on January 1999, mainly at 4.9 GHz. 
The observing frequency was at 4.866 GHz for 16 MHz 
bandwidth with 2-bit sampling. 
The observation was divided into two parts for different 
correlators. From 2000 August 6 UT 21:00 to August 7 UT 02:30, the
participating ground telescopes  were six of the 10 VLBA stations 
(FD, KP, LA, MK, OV, PT), the VLA and ATCA
(Narrabri), with the data correlated at the Soccoro VLBA correlator. 
From August 7 UT 03:00  
to 13:00, the ground telescopes were the ATCA, Ceduna, Mopra, Hartebeesthoek
and VLBA-Mauna Kea, with the data correlated at the Penticton S2 correlator. 
The space data obtained with the HALCA satellite was downloaded and 
recorded in VLBA format at the Green Bank tracking station for 
the first part of the observation, 
and in S2 format at the Madrid, Tidbinbilla and 
Goldstone stations for the latter part.  

The data from the correlators were exported into the Astronomical 
Image Processing System (AIPS) as FITS files, amplitude
calibrated, fringe-fitted, and averaged before standard 
VLBI imaging procedures in Difmap. 
The source was highly resolved on the longer baselines, and fringe-fitting 
for HALCA was only successful for projected baselines up to an 
earth diameter. This was consistent with the previous VSOP series on this 
source (Fujisawa et al. 2000) in which no space fringes were detected. 
The ($u,v$) coverage of the data used for 
imaging is shown in Figure 1.  The image obtained from this ($u,v$) coverage 
is shown in Figure 2. The rms noise level of the space VLBI image is 
0.90 mJy/beam for the peak level of 182 mJy/beam, corresponding to 
a dynamic range 
of 200. The image obtained from the ground baselines only (solid lines 
in Figure 1) is shown in Figure 3. This image yields an rms noise level 
of 0.57 mJy/beam and a peak level of 167 mJy/beam, 
hence a dynamic range of 293. 

\begin{figure}
  \begin{center}
% \FigureFile(180mm,180mm){figure4nn.eps}
 \FigureFile(80mm,80mm){figure4a.ps}
 \FigureFile(80mm,80mm){figure4b.ps}
%\plottwo{figure1.ps}{figure1.ps}
    %%% \FigureFile(width,height){filename}
  \end{center}
  \caption{
(a) left: 5.0 GHz VLBA image at 1999 May 4 (Tingay and Murphy 2001), 
(b) right: Image
from the same data for Figure 3.  
Both images are restored with the larger beam of 19 $\times$ 6 mas at position 
angle 4$^\circ$ as used by Tingay and Murphy (2001) for spectral analysis.
Contour levels are -1, 1, 2, 4, 8, 16, 32 and 64\% of 0.854 Jy/beam, 
the peak flux density in the map, for (b), and
$-$0.5, 0.5, 1, 2, 4, 8, 16, 32 and 64\% of 0.839 Jy/beam for (b).}\label{fig:sample}
\end{figure}

\begin{figure}
  \begin{center}
%    \FigureFile(80mm,80mm){Figure5lb.eps}
    \FigureFile(80mm,80mm){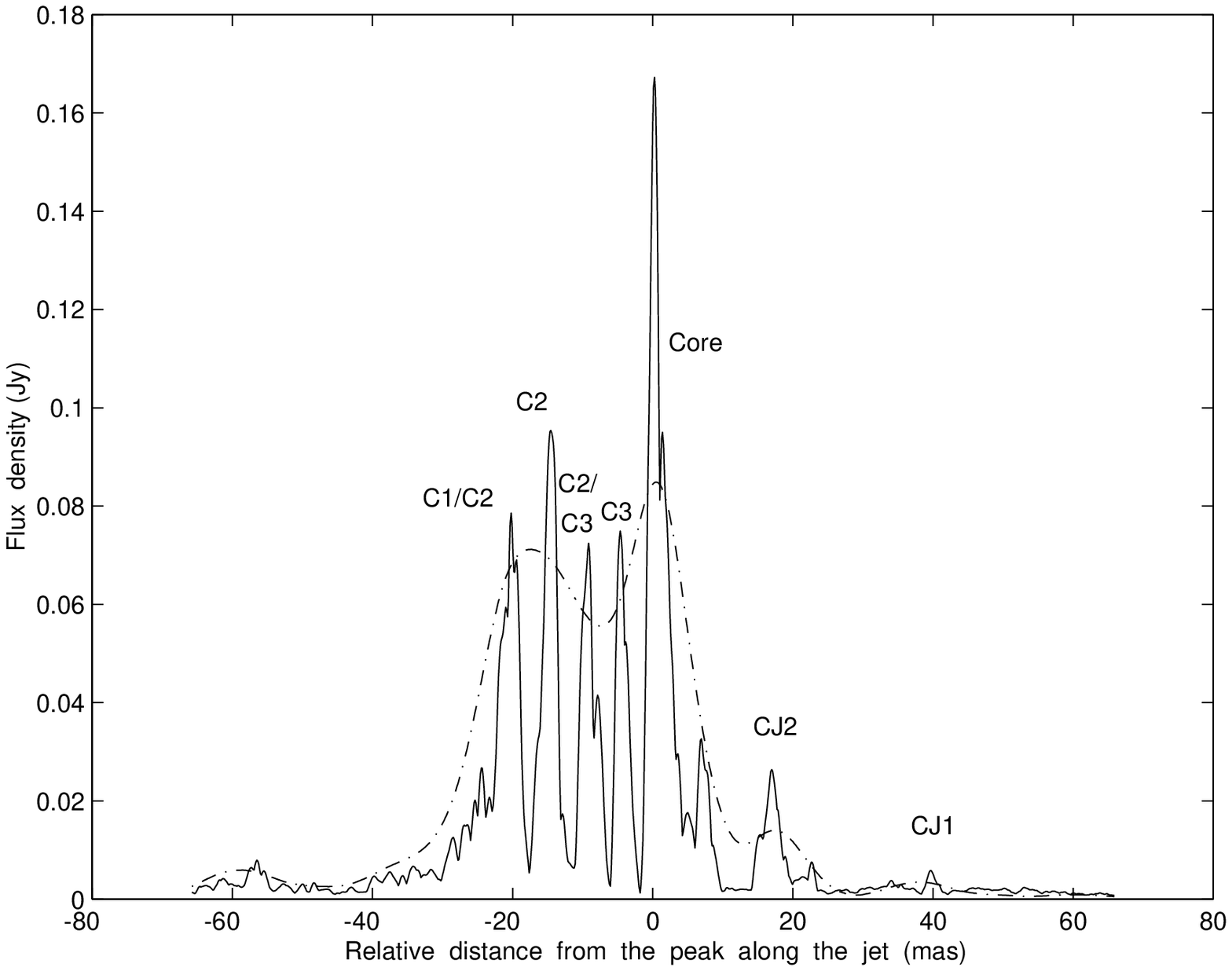}
    %%% \FigureFile(width,height){filename}
  \end{center}
  \caption{Surface brightness profile image along the P.A.\ 51$^\circ$ 
corresponding to the jet axis. The line represents 
the ground baseline image shown in Figure 1 and the dotted line 
represents that convolved with the larger beam shown in Figure 4, 
with the amplitude scaled to 0.1 times for comparison. 
The identification of the components for the Core, C1, C2, C3, CJ1 
and CJ2 are adopted from Tingay, Preston and Jauncey (2001). 

}\label{fig:sample}
\end{figure}

In both images the jet is remarkably linear and oriented along the
position angle $\sim$ 51$^\circ$ with several components visible. The
brightest part of the jet has an elongation northeast and a counter
jet to the southwest.  It is not easy to distinguish between the core
and the jet with a single epoch, single frequency map for such a
complicated source.  The nucleus region of Centaurus A is affected
substantially by both synchrotron self-absorption and free-free
absorption, which are both frequency-dependent effects.  Tingay and
Murphy (2001) registered images of 2.2 GHz, 5.0 GHz and 8.4 GHz 
restored using an average restoring beam of 19 $\times$ 6 mas at
position angle 4$^\circ$, in order to estimate the free-free optical
depth and the intrinsic spectral index.  They found that the core has
a comparable flux to the jet at 5 GHz, while the core dominates the jet
at 8 GHz and the core is substantially absorbed and invisible at 2.2
GHz.  To identify the core region we compared Tingay and Murphy's 5.0
GHz VLBA image at 1999 May 4 (Figure 4-a) and our image (Figure 4-b),
both restored with the same beam as originally used in Tingay and
Murphy (2001).  In spite of the 15 month separation between these two
observations, the images appear to be very similar at this
resolution, hence we concluded that the brightest part around the
phase center in our image corresponds to the core region, in which 
the 2.2
GHz flux would be substantially absorbed as observed by Tingay and
Murphy (2001).

The surface brightness of both the original image (Figure 3) and the lower 
resolution image (Figure 4-b) is illustrated in Figure 5
as a ``slice'' profile along the jet. 
Assuming that the elongated region that includes the brightest part 
is the core, the identification of large scale components over 60 mas, 
C1, C2, C3 CJ1 and CJ2, are adopted from 
comparison to the previous VLBI monitoring (\cite{key-3}, \cite{key-4}). 
In our image there are 4 distinctive bright parts in the jet for 
C1, C2 and C3, which leave some ambiguity in the identification. 
Hence we tentatively denote a component between possible 
C2 and C3 as "C2/C3", and another component in northeast of C2 as "C1/C2". 
%We will discuss this and alternative identification in the next section.  
In the previous VLBI monitoring (\cite{key-3}, \cite{key-4}), both C1 and C2
show elongated linear structure along the jet and the fitted lengths are 
sometimes up to around 20 mas depending on the observing epoch. 
It is likely that our observation
resolves those two regions into more than two components. 

Fujisawa et al. (2000) claimed to have found a large bending of jet
close to the position of component C3 from the VSOP observation
at 5 GHz in January 1998. 
Our observation had a much better ($u,v$) coverage and two times better 
resolution perpendicular to the jet: no such bending is recognizable
in our image.   
 However because of $-43^\circ$ declination of Centaurus A and the 
positional relation between the VLBA and the other southern 
hemisphere ground telescopes, our data still suffer large ($u,v$) holes
over the intermediate resolution range of 50-100 M$\lambda$, 
especially at position angles similar to the jet direction.
  To prove the  reliability of our Centaurus A map as
  shown in Fig.3 we made a couple of tests:
  1) We retained clean components for only the 5 brightest features,
  from the core region through the C1/C2 components in the map, 
  and examined how well they alone fit the data. 
  We observed that about 40\% of the flux on the shortest baselines 
  correspond to the residuals while the visibilities on the longer 
  baselines fit the model very well,
  indicating there is significant structure that is not accounted for in 
  the simple map. Finally we exported the retained clean components to the 
  original uv-data before clean-selfcalibration iteration and confirmed 
  that the residual map clearly showed weaker features such as the counter 
  jet. 2) We tried model-fitting the visibility data and see how many
  components are definitely required before there is ambiguity as to
  what additional components (if any) should be added.
  We found up to 12 required components whose positions are reasonably
  consistent with the features in the original map as shown in Fig.3.

To quantify the structure of the nuclear region of Centaurus A, 
the image obtained using the array of ground telescopes only
(Figure 3) was analyzed 
using the AIPS task JMFIT to modelfit each sub-component with a
Gaussian profile. This procedure is more reliable than fitting models
to the ($u,v$) data when the high-resolution visibilities are too 
complex to fit the model meaningfully. Table 1 shows the parameters 
obtained from the modelfit. 
The brightest core feature around the phase center is modeled with 
4 components labeled with Core1, Core2, Core3 and Core4. 
The brightness temperature T$_b$ is also derived for each component
and listed in Table 1.

\begin{table}
  \caption{Models for the image shown in Figure 3}\label{tab:LTsample}
\begin{center}
 \begin{tabular}{cccccccc}
  \hline\hline
%  S(Jy) & d(mas) & $\theta$(deg) & A(mas) & B/A(mas) & $\psi$($^\circ$) & I.D. \\
% S & d & $\theta$ & A & B/A & $\psi$ & T$_b$ & ID \\
  Flux & Radius & $\theta$ & Major axis & Axial ratio & $\psi$ & T$_b$ & ID \\
  (Jy) & (mas) & (deg) & (mas) &   & (deg) & 10$^{10}$K & \\
  \hline
%0.155 & 0.00 & 0.0 & 1.49 & 2.50 &  52.7 & 2.19 & Core \\
0.478  & 0.00 & 0.00 & 1.52 &   0.49 &  50.6 &2.19 &    Core1 \\
%0.465 & 0.10 & $-$63.4  & 4.72 & 0.18  & 46.2 & 0.62 & Core2 \\
0.235  & 1.98 & $-$139.1 & 2.76 & 0.17  & 53.5 & 0.91 & Core2 \\
0.370  & 4.96 & $-$141.5 & 2.13 & 0.62  & 80.7 & 0.68 & Core3 \\
0.121  & 7.14 & $-$139.0 & 1.75 & 0.63  & 47.6 & 0.32 & Core4 \\
\hline
0.274  & 5.10 & 45.0    & 1.84  & 0.58  & 61.3  & 0.72 & C3 \\
0.326  & 9.39 & 49.3    & 2.59  & 0.32  & 45.9  & 0.78 & C2/C3? \\
0.416  & 15.11& 48.5    & 2.10  & 0.45  & 45.4  & 1.09 & C2 \\
0.403  & 20.79& 49.1    & 2.44  & 0.45  & 44.0  & 0.78  & C1/C2? \\

0.118  & 16.8 & $-$141.5 & 2.17 & 0.55  & 77.8  & 0.24 & CJ2 \\
0.040  & 40.3 & $-$139.7 & 5.79 & 0.17  & 0.65  & 0.04 & CJ1 \\

%..... & .....& & & & & \\
% yyyyy & zzzzz & & & & & \\

\hline
\end{tabular}
\end{center}
\end{table}

\section{Discussion}

\subsection{Identification of the core and the brightness temperature}

We resolved the core region into several components. 
We also successfully detected a few components of the counter jet. 
As we described earlier, it is difficult to identify the core itself
from only one epoch at a single frequency. 
This is especially the case if we have no other images to compare 
at similar resolution. 
If the jet and the counter jet are 
intrinsically symmetric and ejected from the core symmetrically,
we should be able to know the location of the core as the origin 
of both the jet and counter jet. 
Under the simple beaming model for symmetric jet system with 
equal and constant bulk and pattern speed, 
the ratio of the apparent distances from the core of equivalent 
components in the approaching and receding jet, in terms of the 
bulk velocity of material in the jet ($\beta = v/c$) and the angle
to the line of sight $\theta$, is $D = (1 + \beta cos \theta)/
(1-\beta cos \theta) \sim 1.1$ for the typical values of  
$v \sim 0.1c$ and $\theta \sim 60^\circ$ adopted from the monitoring 
results of Tingay et al. (1998) and Tingay, Preston and Jauncey (2001)
for the jet components C1 and C2.  
This gives an upper limit of displacement of the core from 
apparent geometrical center which is only applicable for 
outer jet components such as C1 and C2, and there would be no such 
displacement expected for inner components such as C3 
since no significant motion has been recognized in the monitoring.  
The derived mild upper limit of $D$ for Centaurus A is in contrast with 
other radio source 3C 84 in which a relatively large value of
$D = 1.8$ is derived (Walker, Romney and Benson 1994),
suggesting a larger jet speed ($v \sim 0.33c$) and 
a smaller viewing angle ($\theta \sim 32^\circ$)
for the value of jet-counterjet brightness ratio $R$ they measured. 
If CJ2, located around 20 mas from the core, is the counter part of C2,
then $R \sim$ 4. This is consistent with
the 8\,GHz result discussed by Jones et al.\ (1996) for their beaming model.

The brightness temperature of the brightest component, Core1, 
is $2.2 \times 10^{10}$K while Core2, the second most compact component
in the core region, has a brightness temperature of 
$9.1 \times 10^{9}$K (Table 1). Both the brightness temperatures 
are well below the inverse Compton limit.  
It is not yet clear how the free-free optical depth varies 
around the core at this resolution, but it is likely that the core
region is substantially absorbed, hence much brighter intrinsically. 
Assuming that the values derived by Tingay and Murphy (2001) are still 
applicable at this resolution, we can adopt the optical depth of 0.2 at 5 GHz 
and intrinsic spectral index of 2.0 for the unresolved nuclear component, 
in which case the observed flux is expected 
to have been decreased to $\sim$ 7\% of the original flux,
indicating the intrinsic brightness temperature $\sim 3.1\times10^{11}$K 
and $1.3\times10^{11}$K for Core1 and Core2 respectively. 
These values are still below the inverse Compton limit, 
but close to the limit calculated based on 
equipartition arguments (Readhead 1994). 

Kellermann et al. (1997) measured a core size of 0.5 mas for 
Centaurus A with non-imaging visibility analysis
of a few hours VLBA data taken at 43 GHz in April 1994,
fitting the visibilities with a circular Gaussian component.
They also deduced the lower limit of the brightness temperature 
to be about $10^{10}$K for the peak visibility amplitude of about 5 Jy. 
  The $(u,v)$ distance range of our 5 GHz data is similar to the VLBA
  at 43 GHz and we obtained the same rough source size and
  brightness temperature for our 5 GHz core component with
  ten times weaker flux. Although we measured similar values for the
  core brightness temperature for the 5 GHz core,
  if we consider the optical depth of 0.2 at 5 GHz and the intrinsic
  spectral index of 2.0, an estimated "equivalent" intrinsic flux density
  at 43 GHz for the 5 GHz core is about 8 times larger than the 43 GHz 
  core flux density measured in April 1994. This difference may be 
  attributed to significant variability of the core
  that has been observed at high frequencies such as 22 GHz and 43 GHz
  (Botti and Abraham 1993).

\subsection{Comparison with M87}

Marconi et al. (2001) have estimated the mass of the central black hole in 
Centaurus A to be 2.0$\pm_{1.4}^{3.0} \times 10^8M_\odot$, from 
infrared spectra obtained at the Very Large Telescope. 
The Schwarzchild radii, $r_s$, for this mass is approximately 0.00002 pc,
%Apparent angular size of the black hole at 4.3 Mpc with this mass 
%is approximately 2 $\mu$arcsecond, 
hence the spatial resolution of our observation 0.6 mas
(0.012 pc at 3.4 Mpc distance) corresponds to 600$r_s$. 
Junor et al.\ (1999) presented a 43 GHz VLBI image of the jet in M87 
(at a distance of 14.6 Mpc) on an angular scale of 0.12 mas, 
corresponding to a spatial scale of 0.009 mas, or
30$r_s$ for the estimated black hole mass of 
3 $\times 10^9M_\odot$. Their image shows a remarkably broad jet 
having an opening angle of $\sim 60^\circ$ near the center, with 
strong collimation of the jet occurring at $\sim 30 - 100r_s$ 
from the center, and collimation 
continuing out to $\sim$ 1000$r_s$. 
This seems to favor MHD disk wind models. In a seminal paper 
on this subject, Blandford and Payne (1982) demonstrated by 
simple analysis that ionized gas frozen onto magnetic 
field lines at the surface the accretion disk will be unstable 
to outflow,
if the angle of the field line to the vertical is $\theta \geq 
30^\circ$, in which case the material will be flung outward 
along the field lines by 
centrifugal force. The large opening angle of the jet 
seen at 30$r_s$ distance from the center of M87 seems to 
suggest that the jet originates from such a disk outflow
(Junor et al. 1999, see also Meier et al. 2001 for a review).  
Because of the high linear scale
resolution available to resolve the region around
the black hole, Centaurus A is another ideal candidate 
for studies of the jet collimation mechanism. 

Tingay, Preston and Jauncey (2001) examined their Centaurus A data 
to see if they could measure a jet opening angle on scale greater
than their resolution ($\sim 1000 r_s$) but no evidence for a 
significant jet opening angle could be seen. 
In our image the jet appears to be significantly collinear 
from the core region to 60 mas (1pc).  
If we assume Core1 is the core, the jet opening angle 
of the first jet component C3 is $\sim 12^\circ$ at a distance 
of 0.1 pc, or $\sim 5000r_s$ from the core. 
Interestingly, this angle and $r_s$ distance are very similar 
to the results of the 5 GHz VSOP observations of M87 included 
in Junor et al. (1999, Figure 3) as a part of the plot showing
jet full opening angle as function of distance from the core,
although our distance of C3 from the core corresponds to 0.1 pc
and their distance corresponds to 2 pc from the core. 
Figure 6 shows all the measured jet opening 
angles for the components in the jet, excluding the counter 
jet and the core sub-components. This result indicates that
our observation marginally resolves the lower end of the collimation 
region in the Centaurus A jet. 
Future space VLBI missions such as VSOP-2 at 22 GHz and 43 GHz 
(Hirabayashi et al. 2004) will probably be able to image 
the collimation region, within 1000$r_s$ from the center 
of Centaurus A, together with the accretion disk itself.

\begin{figure}
  \begin{center}
%    \FigureFile(80mm,80mm){dist-angle_revised.eps}
%    \FigureFile(80mm,80mm){dist-angle_revised_kg.ps}
    \FigureFile(80mm,80mm){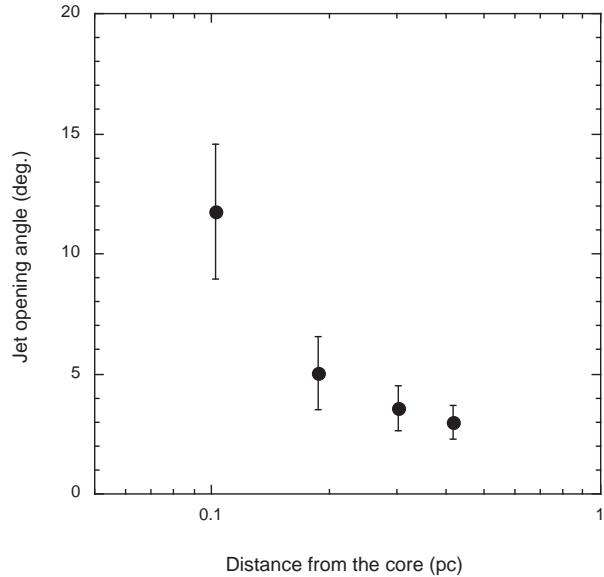}
    %%% \FigureFile(width,height){filename}
  \end{center}
  \caption{Jet full opening angle as function of 
distance from the core for Centaurus A. Error bars 
represent angles corresponding to one fourth of 
beam size to indicate uncertainties of measurement. 

}\label{fig:sample}
\end{figure}

%\newpage

\section{Conclusion}

We have presented a high resolution VLBI image of the nearest radio 
galaxy Centaurus A at 4.9 GHz on 2000 August 6 that shows the sub-parsec 
scale structure of the jet and the core on scales from 0.5 to 
approximately 50 milli-arcseconds from the core 
(0.01--1 pc projected linear distance). The elongated core region is 
resolved into several components over 10 mas long. 
Model-fitting analysis have shown that the strongest peak 
is a compact component with the brightness 
temperature 2.2 $\times 10^{10}$K.
The jet-to-counterjet brightness ratio is approximately 4
for the jet around 20 mas from the core, consistent with
previous observations at 2.3 and 8.4 GHz. 
Assuming the strongest component to be the core,
the jet opening angle at $\sim 5000r_s$ from the core
is estimated to be $\sim 12^\circ$
with collimation of the jet to $\sim 3^\circ$
continuing out to $\sim 20,000r_s$, 
suggesting that collimation of jet occurs inside this region 
similar to that seen in M87.

\vspace{0.5cm}

Part of this work was carried out at the Jet Propulsion Laboratory, 
California Institute of Technology, under contract to NASA,
%Part of the work described in this paper was carried out 
%at the Jet Propulsion Laboratory, Caltech/NASA, 
while S.H. held a National Research Council research associateship.
We would like to thank P. Edwards for helpful comments. 
We gratefully acknowledge the VSOP Project, which is led by 
the Japanese Institute of Space and Astronautical Science 
in cooperation with many organizations and radio telescopes 
around the world. 

%%%%%%%%%%%%%%%%%%%%%%%%%%%%%%%%%%%%%%%

%\begin{table}
%  \caption{This is the first tabular.}\label{tab:first}
% \begin{center}
%    \begin{tabular}{llll}
%      a & b & c & d \\
%      e & f & g & h \\
%      ....
%    \end{tabular}
%  \end{center}
%\end{table}

%\appendix
\section{Conclusions}

%\section{Approximation of ...}

%\section*{Complete data}

%%%
% See the manual for the detail.
%%%

\end{document}